\documentclass[10pt,twocolumn,letterpaper,colorlinks=true, allcolors=blue]{article}

\usepackage[switch]{lineno}

\usepackage{polyglossia}
\setmainlanguage{english}
\usepackage[T1]{fontenc}
\usepackage{float}
\usepackage{enumerate}
\usepackage{xcolor}
\usepackage{graphicx}
\usepackage{amssymb}
\usepackage{orcidlink} 
\usepackage{authblk}
\usepackage{amsmath}
\usepackage{siunitx}
\DeclareSIUnit{\electronvolt}{eV}

\usepackage[
backend=biber,
style=numeric-comp,
sorting=none,
minnames=10,
maxnames=10,
date=year, 
giveninits=true, 
isbn=false,
]{biblatex}

\AtEveryBibitem{
    \clearfield{urlyear}
    \clearfield{eprinttype}
    \clearfield{eprint}
    \clearfield{number}%
    \clearfield{month}%
    \clearfield{issue}
}

\def\graffito@setup{%
   \itshape\footnotesize\leavevmode\color{Black}%
   \parindent=0pt \lineskip=0pt \lineskiplimit=0pt %
   \tolerance=2000 \hyphenpenalty=300 \exhyphenpenalty=300%
   \doublehyphendemerits=100000%
   \finalhyphendemerits=\doublehyphendemerits}

\date{}

\title{Primer of Strong-Field Quantum Electrodynamics \\for Experimentalists}
\author[1]{Annabel Kropf\orcidlink{0009-0006-4500-8462}}

\author[1,2]{Ivo Schulthess\orcidlink{0000-0002-5621-2462}\thanks{Corresponding author: \href{mailto:ivo.schulthess@desy.de}{ivo.schulthess@desy.de}}}

\affil[1]{Deutsches Elektronen-Synchrotron DESY, 22603 Hamburg, Germany}
\affil[2]{Institute for Particle Physics and Astrophysics, ETH Zurich, 8093 Zurich, Switzerland}

\begin{document}
\maketitle

\begin{abstract}
This document serves as a conceptual and practical introduction to Strong-Field Quantum Electrodynamics (SFQED), written from the standpoint of experimental physicists. Rather than providing a comprehensive theoretical review, the document focuses on the core ideas, terminology, and challenges in SFQED that are most relevant to experimental design and interpretation. Our goal is to offer a first point of contact with the subject, bridging the gap between foundational theory and hands-on experimental work, and complementing more formal literature in the field. 
\end{abstract}

Strong-Field Quantum Electrodynamics (SFQED) describes how matter and light behave in electromagnetic fields intense enough to modify the quantum vacuum. Such conditions arise in a variety of physical settings, including laser-particle collisions, plasma wakefield accelerators, ultra-peripheral heavy-ion collisions, and astrophysical environments such as the vicinity of neutron stars and magnetars. In this review, we place particular emphasis on strong-field phenomena that can be explored in controlled scattering experiments, especially laser-particle interactions.

In QED, interactions are computed perturbatively as a series expansion in the coupling constant $\alpha$, making the theory highly predictive at low energies where the coupling remains small. However, in the strong-field regime, this approach can break down because the external field scales up the effective interaction strength, making all orders of the charge--laser interaction important.

This paper introduces the core ideas and challenges of SFQED in an accessible way. It was developed in the context of the \textit{Strong-Field QED Workshop 2024} at DESY, aimed at supporting early-career researchers entering the field from the experimental side~\cite{StrongFieldQED2024}. 
SFQED emerged through sustained collective efforts over several decades. Key foundational contributions include early studies of vacuum nonlinearity by Heisenberg, Euler, Kockel, and Weisskopf~\cite{Heisenberg:1936nmg, Weisskopf:1936hya}, the seminal work of Schwinger on the effective action~\cite{Schwinger:1951nm}, and the exact solutions of the Dirac equation in plane-wave backgrounds by Volkov~\cite{Wolkow:1935zz}. These were unified into a systematic framework for intense-field processes through the Furry interaction picture~\cite{Furry:1951bef} and the subsequent extensive developments by Brown and Kibble, Nikishov, and Ritus~\cite{Brown:1964zzb, Nikishov:1964zza, Ritus:1972ky, Ritus:1985vta}. In particular, the latter’s calculations of nonlinear Compton scattering and electron--positron pair production provided early benchmark results for modern high-intensity laser physics. Complementary to this, the study of vacuum instability in supercritical static fields was pioneered by Greiner and collaborators~\cite{Reinhardt:1977ps, Greiner:1985ce}. While these foundational pillars and recent advancements are covered in several technical reviews~\cite{Marklund:2006my, Blackburn:2019rfv, Fedotov:2022ely, Sarri:2025qng}, the present document is intended as a conceptual primer, bridging the gap between fundamental principles and the specialised literature for those seeking a systematic entry point into the field.

In the following, we examine what characterises a field as strong, the reasons perturbation theory breaks down in such backgrounds, and the conditions under which exact analytical solutions can still be obtained. Central to this discussion are the dimensionless strong-field parameters $\xi$, $\chi$, and $\eta$, which quantify the classical nonlinearity, the quantum nonlinearity, and the energy scale of the interaction, respectively. Finally, we examine how charged particles and photons behave in such fields. We use electrons throughout, but the results are general to other charged particles, particularly positrons. We explore which physical processes, such as nonlinear Compton scattering or nonlinear Breit--Wheeler pair creation, emerge or are modified in this regime. These foundational concepts set the stage for understanding both current and upcoming experiments, such as the Laser Und XFEL Experiment (LUXE) at DESY~\cite{Abramowicz:2021zja, LUXE:2023crk}, and the theoretical tools used to describe strong-field phenomena.

\section{Nonperturbative QED in Strong External Fields}

Quantum Field Theory (QFT) is the theoretical framework for contemporary elementary particle physics, combining classical field theory, quantum mechanics, and special relativity. The Standard Model of Particle Physics is built upon three fundamental QFTs that describe the strong nuclear (Quantum Chromodynamics, or QCD), the weak (Electroweak Theory, or EWT), and the electromagnetic force (Quantum Electrodynamics, or QED), while notably excluding gravity~\cite{Zee:2003mt}. In QFT, particles are treated as quantised excitations of underlying fields, which can interact with one another via the exchange of mediator particles. Often, in QED, it suffices to treat interactions as small perturbations of the underlying free field theory, i.e.\ starting from noninteracting electron and photon fields. In this perturbative QFT approach, these deviations or perturbations are described using series expansions in the coupling constant, which measures the strength of the interaction, in order to approximate observables such as transition amplitudes between particle states.
All traditional perturbative quantum field theories like QED, EWT, QCD, and hence the Standard Model are Lagrangian quantum field theories~\cite{Zee:2003mt, Srednicki:2007qs}. This means the theory can be fully described by a Lagrangian that incorporates the respective symmetries and locality. 

When calculating the probability amplitude for a specific process, the superposition principle dictates that all possible paths or states between the initial and final state must be considered~\cite{Peskin:1995ev}. Each path or interaction history, corresponding to a perturbative contribution, can be visualised using Feynman diagrams, which are based on Feynman rules derived from the Lagrangian. Each vertex in a Feynman diagram introduces another factor of the coupling constant in the perturbative series. When the coupling is less than one, the lowest-order term is dominant, and successive higher-order terms involving more vertices become progressively less significant in the framework of perturbation theory~\cite{Zee:2003mt, Peskin:1995ev}.

Perturbative methods in QED and other quantum field theories can become inadequate for several reasons, two of which are particularly relevant in the context of SFQED.
\begin{enumerate}
    \item Perturbative QFT is based on expanding around the free, noninteracting theory, where interactions are treated as small corrections. This assumes that the physical effects of interactions can be added incrementally. However, in strongly driven systems, nonlinear dynamics can arise where effects feed back and amplify in ways that are not proportional to the initial interaction, requiring a nonperturbative description.
    \item The perturbative approach assumes that higher-order terms in the expansion of observables decrease rapidly enough to ensure convergence. 
\end{enumerate}
Both assumptions can fail in the presence of very strong background fields. More generally, it is important to note that perturbative expansions in QED are not convergent even at weak coupling. As argued by Dyson~\cite{Dyson:1952tj}, the perturbative series in the fine-structure constant is asymptotic, so nonperturbative effects are in principle always present, although they are negligible in weak-field regimes.

When categorising an electromagnetic field as strong or weak, a scale for field strength is implied. QED establishes such a scale using the lowest-mass fundamental charged pair, an electron--positron pair, with the electron mass $m_e$, the elementary charge $e$, and the reduced Compton wavelength $\lambda_c = \frac{\hbar}{m_e c}$, where $\hbar$ is the reduced Planck constant and $c$ the speed of light in vacuum~\cite{Gonoskov:2021hwf}. According to the energy--time uncertainty relation, $\Delta E \Delta t \geq \frac{\hbar}{2}$, the vacuum is subject to constant vacuum fluctuations~\cite{Milonni:1994xx, Mainland:2021ntg}. If the field is strong enough to provide the necessary energy $E = 2 m_e c^2$ for the rest mass of a real electron--positron pair during the fluctuation time $\Delta t$, such pairs can materialise. This process can be understood as field-induced tunnelling, wherein the pair is effectively pulled out of the vacuum, transforming from virtual to real particles. This occurs at the \textit{Schwinger limit}, (Sauter-) Schwinger field, or critical field strength~\cite{Gonoskov:2021hwf}, 
\begin{equation}\label{eq:SchwingerLimit}
    \mathcal{E}_{\text{crit}} = \frac{m_e^2 c^3}{e \hbar} 
                    \approx 1.32 \times 10^{18}~\si{\volt\per\meter} \, .
\end{equation}
For field strengths $|\mathcal{E}| < \mathcal{E}_\text{crit}$ the creation of real particle--antiparticle pairs from the vacuum can still occur via quantum tunnelling, a process that is, however, exponentially suppressed~\cite{Fedotov:2022ely}. The Schwinger limit is thus neither a strict theoretical limit, nor a sharp threshold for pair production, but rather a characteristic field scale for vacuum instability. The physical tunnelling mechanism in strong fields was already identified by Sauter~\cite{Sauter:1931zz} and later incorporated into QED by Heisenberg and Euler~\cite{Heisenberg:1936nmg}. The probability $P$ of electron--positron pair production per unit volume and per unit time in a constant, strong electric field was derived exactly by Julian Schwinger~\cite{Schwinger:1951nm}, extending these earlier results 
\begin{equation}\label{eq:schwingerProbability}
    P = \frac{(e \mathcal{E})^2}{4 \pi^3 \hbar^2 c} \sum_{j=1}^{\infty} \frac{1}{j^2} \exp\left(-\frac{\pi m_e^2 c^3}{e \hbar \mathcal{E}}j\right) \ .
\end{equation}
It is important to note that while the Schwinger limit sets a natural scale for nonperturbative effects in QED, perturbation theory can break down at field strengths well below the critical value $\mathcal{E}_\text{crit}$. This occurs because in interactions involving relativistic particles, the effective field strength in the particle's rest frame is enhanced by the Lorentz factor. As a result, even strong laboratory fields may appear critical to high-energy particles, triggering potentially nonperturbative behaviour before the Schwinger limit is reached.\\

QED is weakly coupled and treated as a power series in the fine-structure constant $\alpha$, which characterises the strength of the electromagnetic interactions. 
This allows for a perturbative approach, since the significance of higher-order terms diminishes, and they can be disregarded for calculations of weak-field QED processes. However, the effective coupling $\alpha$ and therefore the strength of the electromagnetic interaction changes depending on the spatial resolution or energy scale, due to the vacuum becoming increasingly polarised. As a result, $\alpha$ increases with energy and eventually diverges, a threshold known as the \textit{Landau pole}. The Landau pole is predicted within the perturbative framework but lies far beyond both the electroweak unification and Planck scales, rendering it physically inaccessible and of no direct phenomenological relevance within the Standard Model. It is estimated in some models to be $\Lambda_L>10^{227}~\si{\giga\electronvolt}$~\cite{Jian:2020gcm, Gockeler:1997dn}.\footnote{The Landau pole estimation varies with theory, and can be as low as $\Lambda_L \backsimeq 10^{17}~\si{\giga\electronvolt}$ for the Minimal Supersymmetric Standard Model with Four Higgs doublets~\cite{Gockeler:1997dn}. The Planck scale, however, with $1.22 \times 10^{19}~\si{\giga\electronvolt}$~\cite{Zee:2003mt}, lies significantly below most Landau pole estimates. At the Planck, scale traditional concepts, like smooth spacetime, fail. This highlights that QED, as an effective field theory, is expected to break down at energy scales below the Landau pole.} Its appearance instead suggests that perturbative approaches will break down eventually and that QED should be regarded as an effective field theory. The correction to the fine structure constant due to the running of the coupling remains relatively small with only about $7\%$, even at large momentum transfers typical of high-energy interactions~\cite{Davier:1997vd, Jegerlehner:2001ca}. This small correction can therefore be consistently accounted for within the framework of perturbative QED~\cite{Fedotov:2022ely, L3:2005tsb}. 

Another possibility to reach the nonperturbative regime of QED is by scaling up the coupling by introducing a strong external field, such as those produced by lasers.
By increasing this background field, the effective, scaled coupling, referred to as $\xi$ or $a_0$, becomes large enough so that the interaction between field and charge must be accounted for to all orders in perturbation theory. This means the charge--field coupling becomes strong, despite a fine structure constant $\alpha$ that remains small, making it a process that is \textit{nonperturbative at weak coupling}~\cite{Fedotov:2022ely, Abramowicz:2021zja}.

\subsection{Strong-Field QED and the Furry Picture}

Strong-field QED refers to regimes in which external electromagnetic fields are so intense that conventional perturbative QED calculations become invalid or insufficient. However, the term \textit{strong field} refers to a wide variety of field configurations that come close to or exceed the critical field strength $\mathcal{E}_\text{crit}$. Beyond the field strength, the applicability of different SFQED approaches depends on the spacetime structure of the background field. In particular, the field geometry, such as its polarisation, focusing, and gradients, determines which SFQED approximations are applicable~\cite{fields}. This leads to the question: What types of strong-field scenarios are solvable and by what methods? 

One approach in QED is to solve the Dirac or Klein--Gordon equation of motion exactly in the presence of an external field. If an exact solution is available, it describes the dynamics of a particle under the influence of the field to all orders in the coupling and hence in the background field amplitude, without requiring an expansion in $\alpha$, truncation of terms, or diagrammatic summation~\cite{DiPiazza:2011tq, Fedotov:2022ely}.

This applies directly to SFQED as well: if the equation of motion can be solved exactly, nonperturbative effects of the background field can be incorporated at the level of single-particle states, allowing to treat the interactions with the quantised radiation field perturbatively within a QFT framework. However, an immediate caveat arises: exact solutions are only known for specific cases, namely background fields with high symmetry~\cite{Heinzl:2011ur, Fedotov:2022ely}. The most prominent example is a classical plane-wave electromagnetic field, which can be treated as a coherent background and for which exact solutions to the Dirac equation are known. Such fields can be produced by idealised intense laser pulses with waist (focal spot size) of $w_0 \gtrsim 2 \lambda_L$, where $\lambda_L$ is the wavelength of the laser~\cite{Blackburn:2023mlo}. In this setting, the electron motion in the field can be solved exactly, yielding what are known as \textit{Volkov states}: exact solutions to the Dirac equation in a plane-wave background~\cite{Wolkow:1935zz,Fedotov:2022ely}. 

The coherent field modifies the motion of the electron in a predictable and smooth way, effectively \textit{dressing} the particle.\footnote{The particle interacts continuously with the field, which modifies its properties (momentum, phase, energy, etc.) These modifications are not momentary but built into the particle’s wavefunction itself, as captured by the Volkov states.}

\begin{figure}[tbh!]
    \centering
    \includegraphics[width=0.6\linewidth]{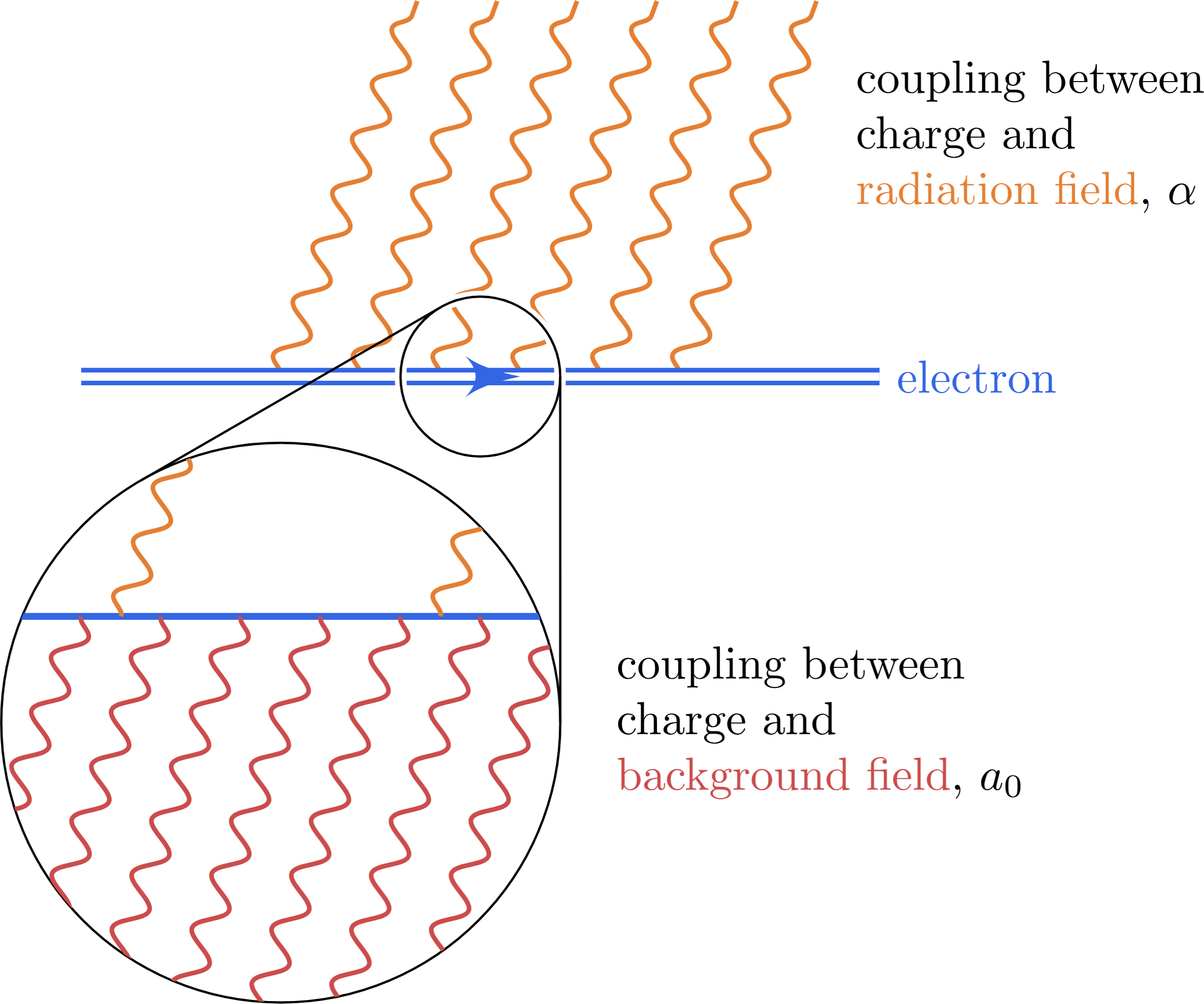}
    \caption{Nonperturbativity arises in two distinct ways: from the coupling between the electron and the strong classical background field, depicted in red, and the coupling between the charge and the quantised radiation field, depicted in orange. The former corresponds to nonperturbative dressing by the background field and can be treated exactly within the Furry formalism by incorporating the background into the particle states. This is distinct from infrared dressing by soft photons, which occurs even in weak-field QED and is handled via infrared resummation techniques such as the Bloch--Nordsieck or Kinoshita--Lee--Nauenberg formalisms~\cite{Bloch:1937pw, Kinoshita:1962ur, Lee:1964is, Weinberg:1995mt}. Figure from~\cite{Gonoskov:2021hwf}. }
    \label{fig:coupling}
\end{figure}

In SFQED, this formalism is known as the \textit{Furry picture}~\cite{Fedotov:2022ely}. Unlike in perturbative QED, where particles are treated as free, the Furry picture uses dressed states (i.e.\ Volkov states) for external legs of Feynman diagrams. Interactions with the quantised radiation field (i.e.\ the emitted real photons) are still treated perturbatively, but the interaction with the strong classical background is incorporated exactly, i.e., nonperturbatively, from the outset, see Figure~\ref{fig:coupling}. This approach is not viable if the interaction becomes nonperturbative in the coupling to the quantised radiation field.

\begin{figure}[tbh]
    \includegraphics[width=\linewidth]{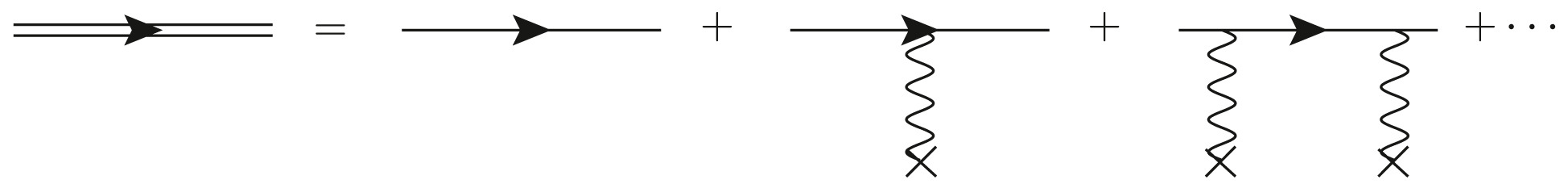}
    \caption{Diagrammatic expansion of the dressed Furry picture propagator, represented with a double line, regarding position-space Feynman diagrams, from~\cite{Fedotov:2022ely}. Each wavy photon line ending with a cross denotes an interaction with a prescribed classical background field via the vertex $-i e \gamma \cdot A_\text{ext}$. The dressed propagator can be viewed as a resummation of an infinite series of such background-field interactions.}
    \label{fig:Furry}
\end{figure}

In Feynman diagrams, dressed particle lines, i.e., particles that continuously interact with the coherent background, are conventionally represented by double lines, and crosses on photon lines are used to indicate interactions with a prescribed classical background field as shown in Figure~\ref{fig:Furry}. 

In the presence of a strong external field, the quantum states of electrons and their evolution in time and space are modified to inherently include field effects, ultimately changing various properties of the particle. Consequently, predictions are influenced by the spatio-temporal characteristics of the field, such as its pulse shape and duration in the case of a laser-induced background field~\cite{Gonoskov:2021hwf}. 
In the laboratory, however, laser fields are seldom fully plane-wave or coherent. To make use of the Furry picture for real-world applications, two approximations may be made depending on the intensity of the background field:
\begin{enumerate}
    \item \textbf{Locally Monochromatic Approximation (LMA)}: While globally, a real laser may be a pulse or a focused beam, it may be approximated as locally behaving like a clean, single-frequency monochromatic plane wave (i.e.\ a wave with a single frequency and infinite extent). This approximation allows the field to retain an oscillatory structure, with parameters such as amplitude and frequency slowly varying in spacetime. Exact QED results derived in idealised monochromatic backgrounds enable processes to be modelled in realistic laser fields, especially when harmonic structure or interference effects are important~\cite{DiPiazza:2011tq, Fedotov:2022ely}. Since the LMA relies on the notion of a local frequency, it is defined for backgrounds that are (locally) plane-wave-like. In practice, it is used for regions where $\xi \sim 1$ since it becomes computationally inefficient for high values of $\xi$~\cite{Blackburn:2023mlo}.
    \item \textbf{Locally Constant Field Approximation (LCFA):} A real laser field may vary in space and time, but zoomed in to small spacetime regions, the field looks approximately constant, both in strength and direction. Thus, this approach is justified when the formation length, i.e., the spacetime region over which the QED process occurs, is small compared to the scale over which the field varies. The approximation allows for using known SFQED results for constant fields locally, like Schwinger pair production or nonlinear Compton scattering rates, and integrating them over the varying background~\cite{Heinzl:2020ynb, Fedotov:2022ely}. This approximation is valid for $\xi \gg 1$ and applied to a variety of strong fields~\cite{Blackburn:2023mlo}. In the plane-wave background, the LCFA can be obtained as the high-intensity limit ($\xi \gg 1$) of the LMA~\cite{Heinzl:2020ynb}. However, it is known to break down in the infrared region, where the formation length of soft processes becomes large compared to the field variation scale.
\end{enumerate}
In the following, without loss of generality, we assume strong, coherent fields produced by a laser.

\subsection{Physical Regimes and SFQED Parameters \texorpdfstring{$\xi, \chi$ and $\eta$}{xi, chi and eta}}

In the presence of strong, coherent electromagnetic fields, electrons exhibit nonlinear dynamics that differs from their behaviour in weak-field or vacuum conditions. These interactions are described using the Furry picture, in which the interaction with a background field is treated exactly. But what determines how a particle behaves in such a field? Two key factors are: (1) the strength and frequency of the background field, and (2) the energy or velocity of the particle traversing it.\footnote{This can be understood intuitively: A particle that is "surfing" a coherent electromagnetic wave experiences a force that causes it to wiggle; the strength of this motion depends on the field amplitude and frequency. If the particle is moving close to the speed of light, it sees the field compressed in its rest frame due to the Lorentz transformation, which further enhances the effective field strength it experiences.} These combine to define three dimensionless fundamental parameters that characterise the regime of SFQED and govern the dynamics of particle-laser interactions, namely~\cite{Abramowicz:2021zja}:
\begin{enumerate}
    \item The classical nonlinearity parameter or intensity parameter, $\xi$ (commonly denoted $a_0$ in laser physics).
    \item The quantum nonlinearity parameter, $\chi$.
    \item The energy parameter, $\eta$.
\end{enumerate}
The classical nonlinearity parameter $\xi$, which was previously introduced as the effective coupling, characterises the strength of the interaction between an electron and a background field. It is defined as~\cite{Heinzl:2008rh, Gonoskov:2021hwf},
\begin{equation}\label{eq:xi}
    \begin{aligned}
        \xi &= \left( e^2 \left< -(p \cdot \mathcal{F})^2 \right> \, / \, \left[m_e k \cdot p \right]^2 \right)^{1/2} \\
        &= |e|\mathcal{E}\frac{\lambda_c}{\hbar\omega_L} =
        \frac{|e|\mathcal{E}}{m_e c \omega_L} \, ,
    \end{aligned}
\end{equation}
where $\mathcal{F}$ is the classical field strength tensor, $p$ is the momentum of the electron, $k$ is the wavevector of the laser field, ${\left< \cdot \right>}$ indicates the phase cycle average, $\mathcal{E}$ is the background field strength, and $\omega_L$ is the angular frequency of the photon.

Physically, $\xi$ may be interpreted as the work performed by the background field over the Compton wavelength of the electron, in units of the photon energy. When squared, $\xi^2$ is proportional to the number of photons from the coherent background field effectively involved in a leading-order process such that an increase in $\xi$ corresponds to an increase in the effective photon number $n_*$.\footnote{This is not a discrete photon count but an effective number of photons from the coherent background field.} This makes it a useful measure of the degree of classical nonlinearity (i.e.\ how many effective photons interact with the particle) in laser-particle interactions. Importantly, $\xi$ is a purely classical parameter, see equation~(\ref{eq:xi}), as it remains meaningful in the formal limit ${\hbar \to 0}$, where quantum effects vanish. It can also be written related to $\mathcal{E}_\text{crit}$ as
\begin{equation}
    \xi = \frac{m_e c^2}{\hbar \omega_L}\frac{\mathcal{E}}{\mathcal{E}_{\text{crit}}} \, .
\end{equation}
In this form, it becomes apparent that $\xi < 1$ means that the process can be described using perturbative QED, while for ${\xi \sim \mathcal{O}(1)}$ the perturbative approach is not valid~\cite{Abramowicz:2021zja, LUXE:2023crk}.

The quantum nonlinearity parameter $\chi$ can be written as 
\begin{equation}
    \chi = \frac{e \hbar \left[-(p \cdot \mathcal{F})^2\right]^{1/2}}{m^3 c^4} \, .
\end{equation}
It quantifies the importance of quantum effects, specifically the degree to which the particle couples to the quantised radiation field in the presence of a classical background. Since both charged particles ($e^\pm$) and neutral photons ($\gamma$) can undergo quantum processes, though in different ways, two distinct definitions are required~\cite{DiPiazza:2011tq, LUXE:2023crk}:\footnote{In contrast, $\xi$ defines how charged particles respond to the classical field.} 
\begin{align}
    \chi_e &\approx \gamma\frac{\mathcal{E}}{\mathcal{E}_\text{crit}}(1+\beta\cos(\theta))\\
    \chi_\gamma &\approx \frac{\hbar\omega_\gamma}{m_ec^2}\frac{\mathcal{E}}{\mathcal{E}_\text{crit}}(1+\cos(\theta))
\end{align}
Here, $m_e c^2$ is the electron rest energy, $\hbar\omega_\gamma$ is the photon energy, and $\theta$ is the angle between the particle velocity and the opposite of the field propagation direction. With this convention, a head-on collision corresponds to $\theta=0$. Both $\chi=\chi_{e,\gamma}$ parameters increase with the strength of the background field and the energy of the particle involved. 

For $e^{\pm}$, $\chi_e$ signals the onset of quantum radiation effects.
In classical electrodynamics, radiation is emitted continuously and carries away only a small fraction of the particle’s energy. In contrast, for $\chi_e \gtrsim 1$, high-energy photons are emitted stochastically, a phenomenon known as \textit{quantum recoil}~\cite{DiPiazza:2011tq, Blackburn:2019rfv}. 

For photons, $\chi_\gamma \gtrsim 1$ indicates that a single high-energy photon in a strong background field can create a real electron--positron pair, meaning the onset of conversion of light into matter.

Physically, $\chi_e$ may be interpreted as the energy transferred from the laser pulse to an electron over a reduced electron Compton wavelength, in units of the particle's rest energy. Alternatively, $\chi_e$ has a geometric interpretation related to the curvature of the particle’s trajectory in spacetime coordinates, its \textit{worldline}. Indeed, the parameter $\chi_e$ can be directly linked to the radius of curvature, $R$, of the particle's worldline in spacetime~\cite{Fedotov:2022ely},
\begin{equation}
    \chi_e=\frac{\lambda_c}{R} \, .
\end{equation}
In SFQED, the extent to which the worldline is bent by the field determines whether a classical description is sufficient or a quantum description is needed. Thus, quantum effects become significant when the field bends the particle’s trajectory on length scales comparable to or smaller than the particle's own Compton wavelength, $\lambda_c$, that is, when $\lambda_c \gtrsim R$, classical descriptions break down, and quantum effects dominate. Notably, if $\chi>1$, perturbative methods still hold in most cases, given that higher-order corrections remain small. It is only when non-negligible loop corrections are present, which happens at $ \alpha\chi^{2/3}\geq1$, that perturbative theory including the Furry expansion is expected to break down despite $\alpha \ll 1$, a prediction referred to as the \textit{Ritus--Narozhny conjecture}~\cite{Fedotov:2022ely}.

The parameters $\chi$ and $\xi$ together characterise the transition from the perturbative to the nonperturbative regimes in SFQED, either through increasing classical nonlinearity (via $\xi$) or growing quantum effects (via $\chi$). They are related via the dimensionless energy parameter $\eta$,
\begin{equation}
    \eta = \frac{\chi}{\xi} \, .
\end{equation}
$\eta$ is the centre-of-mass energy squared in units of the pair rest energy $2m_ec^2$.

\begin{figure}[tbh]
    \centering
    \includegraphics[width=0.7\linewidth]{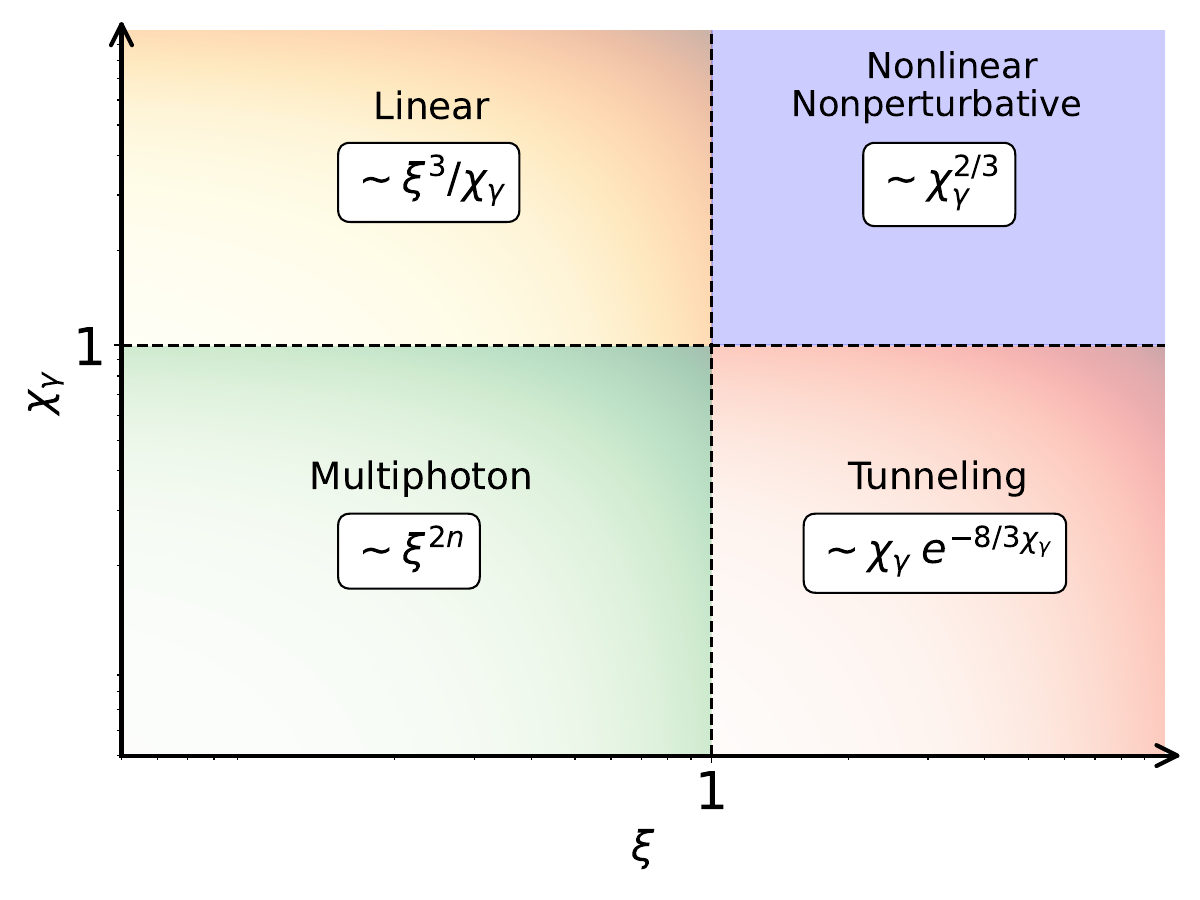}
    \caption{Regimes and related rates for pair production defined by $\chi_\gamma$ and $\xi$~\cite{Abramowicz:2021zja}.}
    \label{fig:regimes}
\end{figure}

When considering a photon in a strong background field, the relevant quantum parameter is $\chi_\gamma$. The resulting interaction regimes are mapped in Figure~\ref{fig:regimes} in the $\xi-\chi_\gamma$ plane.

In the regime $\xi \ll 1$ and $\chi_\gamma \ll 1$, the interaction is already purely quantum, since pair production does not exist classically, but remains multiphoton and perturbative. While individual background photons are too weak to induce pair creation, only channels involving a sufficient number of background-field quanta can satisfy energy--momentum conservation. The process is described by a sum over Feynman diagrams involving an increasing photon number, starting from the minimum kinematically allowed order, with higher-order contributions being increasingly suppressed in the perturbative regime. 

For $\xi \ll 1$ but $\chi_\gamma \geq 1$, the interaction remains perturbative but becomes dominated by strong quantum effects. A photon can be converted into an electron--positron pair through field-assisted decay, typically via a single vertex in the Feynman diagram expansion. The process rate scales linearly with the field intensity, hence the label linear quantum regime.

When $\xi \gg 1$ but $\chi_\gamma \ll 1$, the field is strong, yet the photon energy is insufficient for pair creation. However, the strong field modifies the vacuum structure allowing the photon to tunnel through the energy barrier, with a rate that is exponentially suppressed in $1/\chi_\gamma$. This process is nonperturbative and the background has to be treated exactly. 

Finally, when both $\chi$ and $\xi$ are large ${(\gtrsim 1)}$, the system enters a regime of fully quantum, fully nonlinear SFQED. Here, the Furry picture still holds as long as the background field is coherent and highly symmetric. However, at even higher values of $\chi$, perturbation theory in the number of quantised radiation field interactions may start to break down when $\alpha \chi^{2/3}\geq 1$. Loop corrections become large and the interaction is expected to become fully nonperturbative.

\section{Nonlinear Processes}

With the relevant strong-field parameters established, the distinct processes undergone by electrons, positrons, and photons in strong electromagnetic backgrounds are now examined.\\
Within QED in a plane-wave background, there are four distinct first-order processes: nonlinear Compton scattering, nonlinear Breit--Wheeler pair production, i.e. inelastic photon--photon scattering, and their time-inverse processes, pair annihilation, and photon absorption. They are field-induced, meaning that their probabilities vanish for $\xi \to 0$.
Higher-order processes, such as the nonlinear trident process or elastic photon--photon scattering are also relevant in SFQED, however, much less detailed calculations have been performed due to their complexity in QED even without a background field~\cite{Fedotov:2022ely}.

\subsection{Nonlinear Compton Scattering}

\begin{figure}[tbh]
    \centering
    \includegraphics[width=0.7\linewidth]{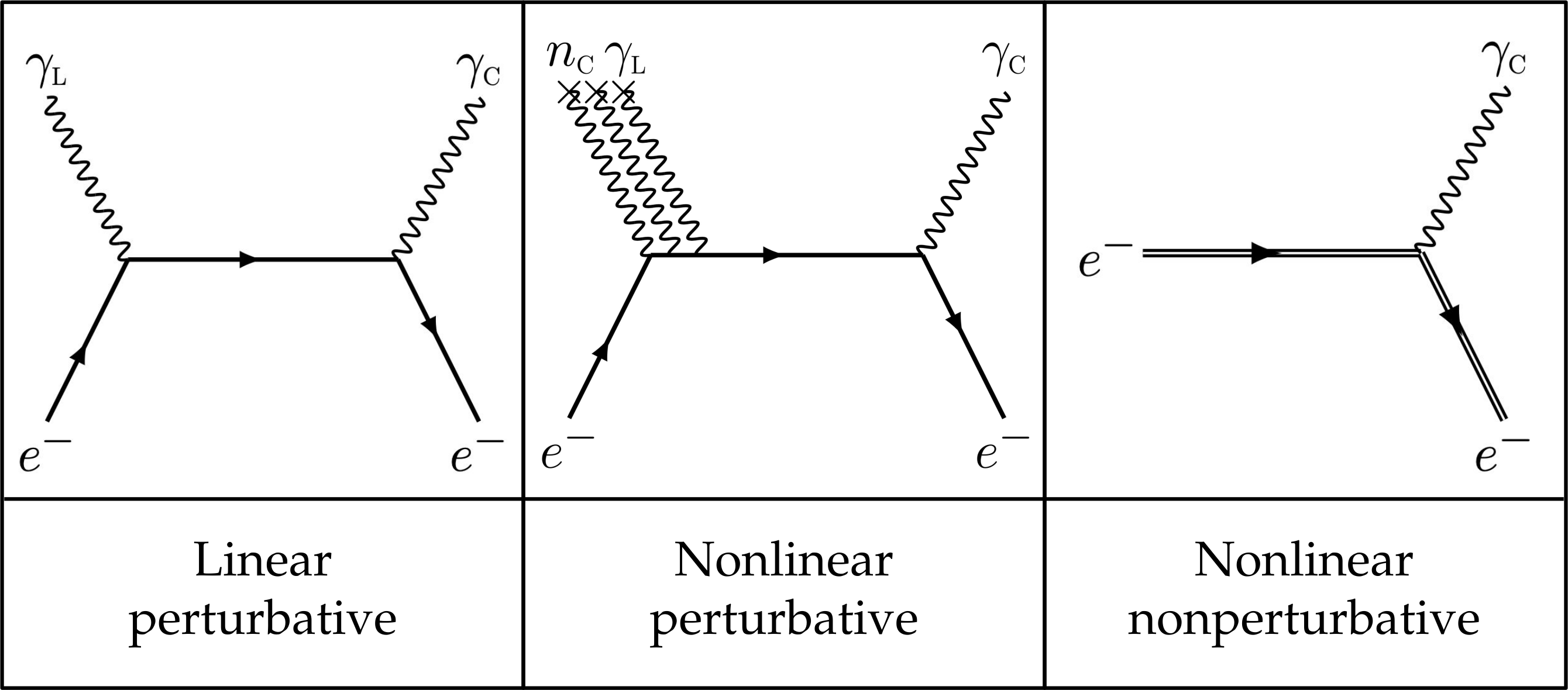}
    \caption{Feynman diagrams for linear, nonlinear perturbative, and nonlinear nonperturbative Compton Scattering. Crosses indicate interactions with the external background field. In the nonperturbative case, dressed electrons are represented by a double line.}
    \label{fig:compton_feynman}
\end{figure}

Linear Compton scattering, $e^- + \gamma_L \to e^- + \gamma_C$, presented in Figure~\ref{fig:compton_feynman} describes the process of one photon colliding with a free or loosely bound electron. As a result, energy and momentum are transferred and the photon is scattered, an interaction leading to a change in wavelength of the photon, as well as a change in the electron's kinematic energy. The Compton process becomes nonlinear when the interaction of the electron with the coherent background field involves an effective photon number $n_*$ and proceeds via higher harmonic channels, producing one high energy Compton photon, $e^-+ n_C \, \gamma_L \to e^- + \gamma_C$, where $n_C \in \mathbb{N}$ labels the harmonic channel.\footnote{We talk in terms of effective photon number $n_*$, because even though we consider a classical field and not distinct photons, the interaction still shows harmonic structure analogous to multiphoton processes and energy and momentum are absorbed in multiples of $\hbar \omega_L$ from the background field.} 
The energy that is transferred depends on the photon's scattering angle.  
The Compton edge represents the highest (lowest) energy that a photon (electron) can have after being scattered, corresponding to the maximum energy transfer scenario where the photon is backscattered~\cite{Abramowicz:2021zja, Fedotov:2022ely, Sarri:2025qng}. It therefore appears as a sharp cut-off in the scattered photon’s (electron’s) energy spectrum, since transferring energy beyond this edge exceeds what is kinematically possible in Compton scattering. In nonlinear Compton scattering, multiple Compton edges appear in the photon (electron) spectrum, reflecting the contribution of higher harmonic channels of the interaction with the background field. The $n$th Compton edge then represents the maximum energy transfer for scattering events associated with the $n$th harmonic of the interaction with the coherent background field.

When the charge--field coupling is small, thus QED is treated perturbatively, the probability of producing a scattered Compton photon is proportional to $\xi^2$ for linear Compton scattering, and $\xi^{2 n_C}$ for nonlinear Compton scattering.  
However, for strong charge--field couplings with $\xi$ exceeding unity, higher-order contributions have to be considered. Due to the increasing interactions with the field, the effective mass of the electron increases with laser intensity,
\begin{equation}\label{eq:emass_shift}
    m_e^{\text{eff}}=m_e^\text{rest}\sqrt{1+\xi^2}.
\end{equation}
Because the Compton edge calculation is based on the electron's mass, it shifts with $\xi$~\cite{Abramowicz:2021zja,LUXE:2023crk}. 
The fractional energy $u$, which represents the photon energy relative to the incident electron energy at the Compton edge, varies across nonlinear QED, nonlinear classical electrodynamics, and linear QED as follows~\cite{Abramowicz:2021zja,LUXE:2023crk}:
\begin{align}\label{eq:CE_shift}
    u_{\text{nonlin.QED}} &= \frac{2\eta}{2\eta + 1 + \xi^2} \, ; \\
    u_{\text{nonlin.class.}} &= \frac{2\eta}{1 + \xi^2} \, ; \\
    u_{\text{lin.QED}} &= \frac{2\eta}{2\eta + 1} \, .
\end{align}
These distinct functional dependencies of $u$ on $\eta$ and $\xi$ provide a means to experimentally discriminate between theoretical models via precise measurements of the Compton edge for the photon spectra.\footnote{Since energy must be conserved, the Compton edge for electron spectra can be calculated using $1-u$.}

\subsection{Photon--Photon Interactions}

Photons can effectively interact with other photons via fluctuating virtual $e^{+}e^{-}$ pairs, a process that is predicted by quantum electrodynamics and would not be possible in classical theories of electromagnetism. 

\begin{figure}[tbh]
    \centering
    \includegraphics[width=0.7\linewidth]{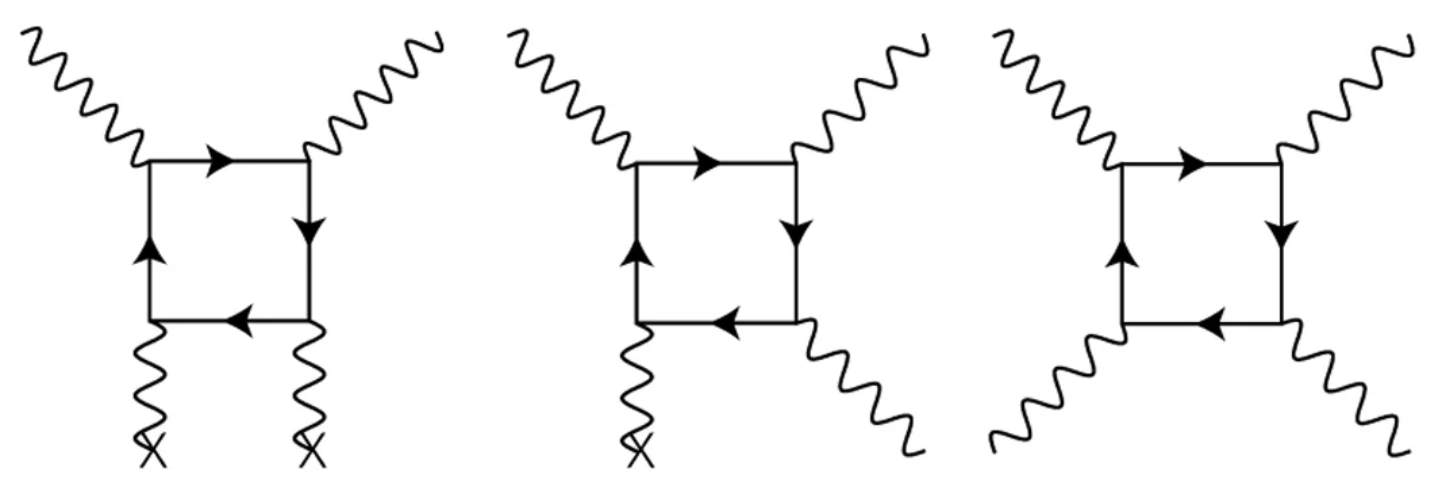}
    \caption{Lowest order virtual one-loop box diagrams for elastic $\gamma\gamma \to \gamma \gamma$ scattering involving fermions, with each cross denoting an external field leg. Despite it being challenging to observe, given that it is an $\alpha^4 \approx 3 \times 10^{-9}$ process, it has been achieved in various experiments. Figure from~\cite{ATLAS:2017fur}.}
    \label{fig:gammagamma}
\end{figure}

Two photons can interact elastically, a process also referred to as light-by-light scattering. The lowest-order diagrams are presented in Figure~\ref{fig:gammagamma}. Vacuum birefringence, wherein the vacuum acts like a birefringence medium altering the polarisation states of traversing light, is, for example, a manifestation of elastic polarised photon-photon scattering~\cite{Borysov:2022cwc}.

\begin{figure}[tbh]
    \centering
    \includegraphics[width=0.7\linewidth]{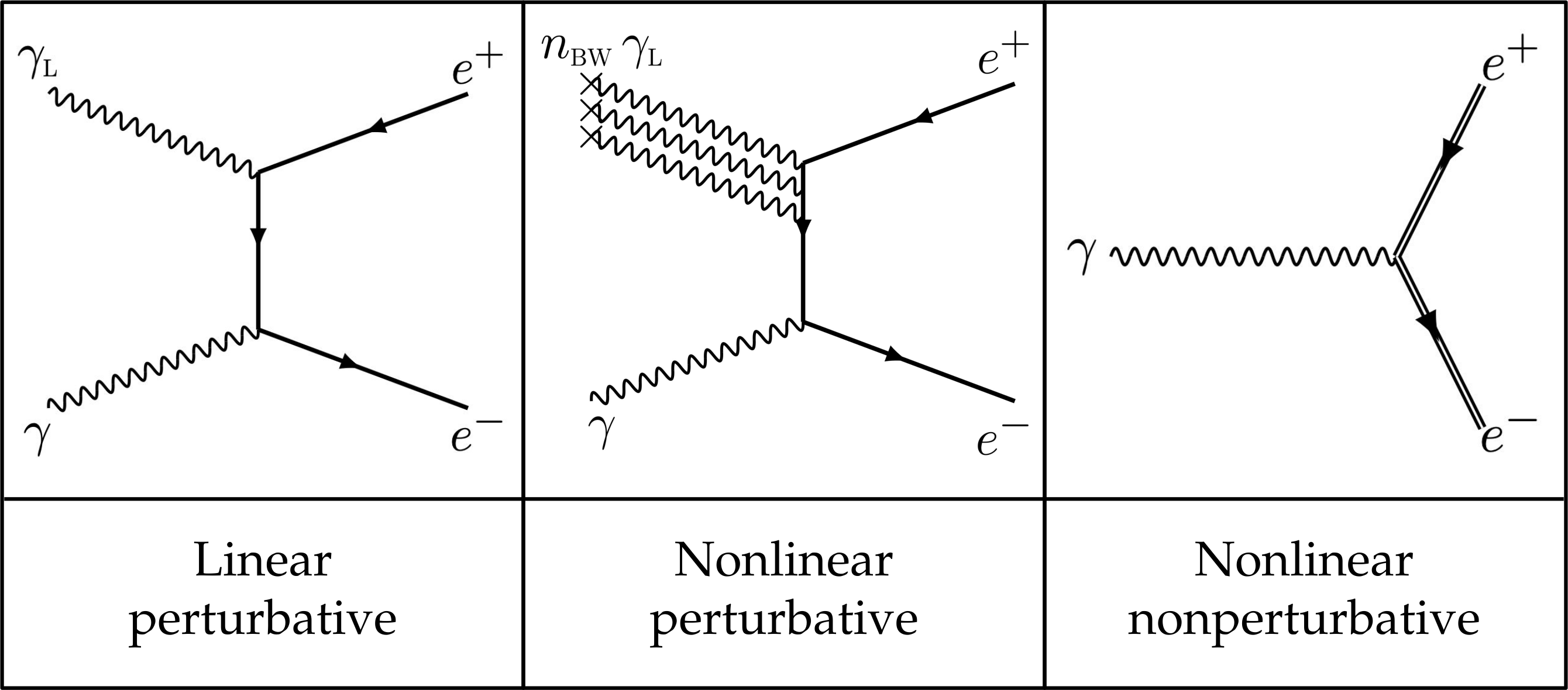}
    \caption{Feynman diagrams for linear, nonlinear perturbative, and nonlinear nonperturbative Breit--Wheeler pair creation. Crosses indicate interactions with the external background field. In the nonperturbative case, dressed electrons are represented by a double line. }
    \label{fig:BW_feynman}
\end{figure}

Inelastic photon-photon scattering leads to the creation of a real $e^{+}e^{-}$ pair, a process called Breit--Wheeler pair creation, presented in Figure~\ref{fig:BW_feynman}. In this case, a real photon absorbs one photon (linear) or multiple photons (nonlinear) and produces a physical $e^{+}e^{-}$ pair, $\gamma + n_\text{BW} \, \gamma_L \to e^{+}e^{-}$~\cite{Abramowicz:2021zja, Hartin:2018sha, King:2024ffy}. In perturbative predictions for low-$\xi$, the Breit--Wheeler rate follows the power-law $\xi^{2 n_\text{BW}}$.

Contrarily, in the tunnelling limit, i.e., $\xi \gtrsim 1/\sqrt{\chi_\gamma} \gg 1$, the rate is given as
\begin{equation}\label{eq:BW}
    \Gamma_\text{BW} \propto \left( \frac{\mathcal{E}}{\mathcal{E}_\text{crit}} \right) \exp \left[ -\frac{8}{3 \gamma_e (1 + \cos \theta)} \frac{\mathcal{E}_\text{crit}}{\mathcal{E}} \right] \, .
\end{equation}
The effective energy available for pair production depends on the photon's energy and its collision geometry with respect to the laser field. This orientation introduces an angle dependence, where the angle $\theta$ represents the relative orientation between the photon's propagation direction and the opposite of the laser propagation direction.
The Breit--Wheeler rate resembles the leading ($j=1$) contribution to the Schwinger rate in equation~(\ref{eq:schwingerProbability}) calculated for a static field $\mathcal{E}_S$~\cite{Hartin:2018sha, Abramowicz:2021zja},
\begin{equation}
  \Gamma_{\text{Schwinger}} \propto \left( \frac{\mathcal{E}_S}{\mathcal{E}_{\text{crit}}} \right)^2 \exp \left[ -\pi \frac{\mathcal{E}_{\text{crit}}}{\mathcal{E}_S} \right] \, .
\end{equation}

When a high-energy electron beam interacts with a high-intensity laser, the pair production may happen as a two-step trident process through nonlinear Compton and subsequent nonlinear Breit--Wheeler, where 
$$e^- + n_C \, \gamma_L \to e^- + \gamma_C \text{~and~} \gamma_C + n_\text{BW} \, \gamma_L \to e^+e^- \, ,$$
or directly as in one-step 
$$e^- + n_\text{tri} \gamma_L \to e^- + e^+e^- \, .$$
Generally, the two-step process dominates at high intensity. At very low intensity, the two-step channel effectively closes and the one-step process dominates~\cite{King:2013osa, Dinu:2019wdw}.

\section{SFQED Environments}

There are a few relevant environments where strong fields are found: The magnetic fields surrounding compact astrophysical objects such as magnetars and black holes, the boosted Coulomb field of dense lepton bunches, the coherently summed nuclear electric fields experienced by leptons travelling through a crystal along a symmetry axis, the electromagnetic fields generated by the focusing of high-power lasers~\cite{Gonoskov:2021hwf}, and
the Coulomb field around nuclei with exceptionally high atomic numbers above $Z\simeq 80$~\cite{Abramowicz:2021zja}.

\subsection{Laser-Particle Experiments}

Lasers allow for the creation of controllable and well-defined ultra-strong electromagnetic fields that can be propagated over macroscopic distances and can persist for extended periods with focal intensities as high as $10^{22-23}~\mathrm{W\,cm^{-2}}$ using chirped-pulse amplification. Generally, experiments aim to reach strong fields by colliding laser--laser~\cite{Grismayer:2016gae}, laser--X-ray~\cite{Ahmadiniaz:2024xob}, laser--plasma~\cite{Slade-Lowther:2018kgv}, or laser--electron-beam~\cite{Bamber:1999zt, reisE320ProgressFY242024, Athanassiadis:2025git, Abramowicz:2021zja, LUXE:2023crk}. In the latter two cases, the laser acts as an accelerator and target~\cite{Gonoskov:2021hwf}. 
The peak intensity $I_0$ of a laser can be directly related to $\xi$ and $\chi_e$, 
\begin{align}
    \xi &\simeq 26.9 \, I_0^{1/2}[10^{21}~\si{\watt\per\centi\meter\squared}] \,\lambda_L[\si{\micro\meter}], \\
    \chi_e &\simeq 0.18 \, E_0[\si{\giga\electronvolt}] \, I_0^{1/2} [10^{21}~\si{\watt\per\centi\meter\squared}],
\end{align}
where $\lambda_L$ is the laser wavelength and $E_0$ the electron beam's initial energy. In laser--electron beam experiments, high-intensity lasers are combined with electrons from an accelerator, such as in the experiments LUXE~\cite{Abramowicz:2021zja,LUXE:2023crk}, SLAC E-144~\cite{Burke:1997ew}, and E-320~\cite{reisE320ProgressFY242024, Athanassiadis:2025git} or, for all the others, electrons produced by laser wakefield acceleration~\cite{Streeter:2020elr, Poder:2017dpw, Cole:2017zca}. An electron beam can also be converted to high-energy photons, which then collide with the laser. 
In these setups, typical signatures of QED effects are investigated, such as the influence of the electron’s effective mass on the harmonics in radiation spectra and on the energy of scattered photons and electrons in nonlinear Compton scattering. Another key observable is the rate dependence on $\xi$ in nonlinear Breit--Wheeler pair creation.

\begin{figure}[tbh]
    \centering
    \includegraphics[width=0.7\linewidth]{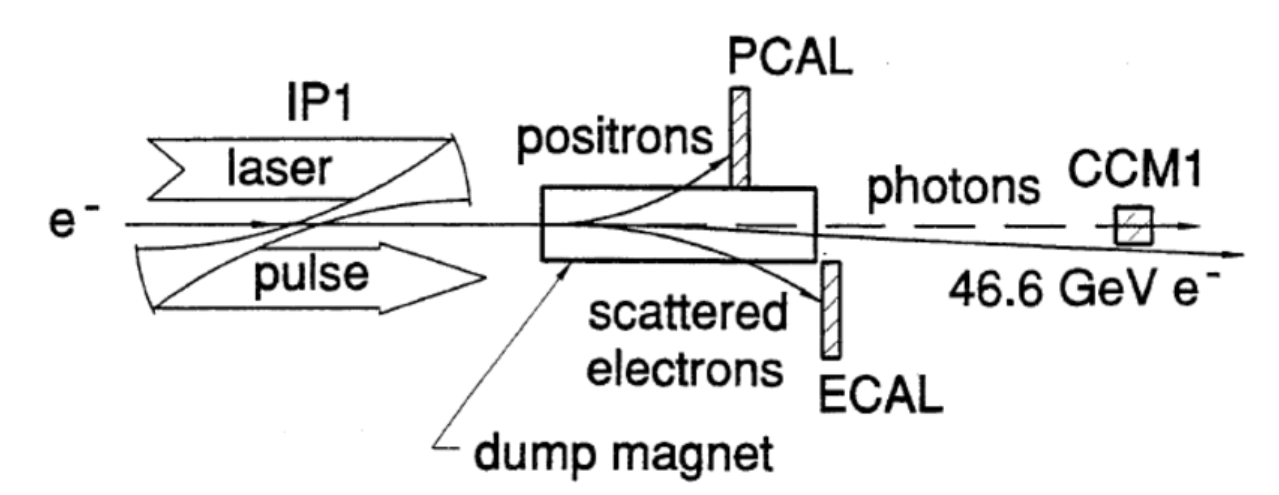}
    \caption{Collision region layout of the pioneering E-144 experiment conducted at SLAC, from~\cite{Burke:1997ew}. }
    \label{fig:E-144}
\end{figure}

\begin{figure*}[tbh]
\centering
\includegraphics[width=1\linewidth]{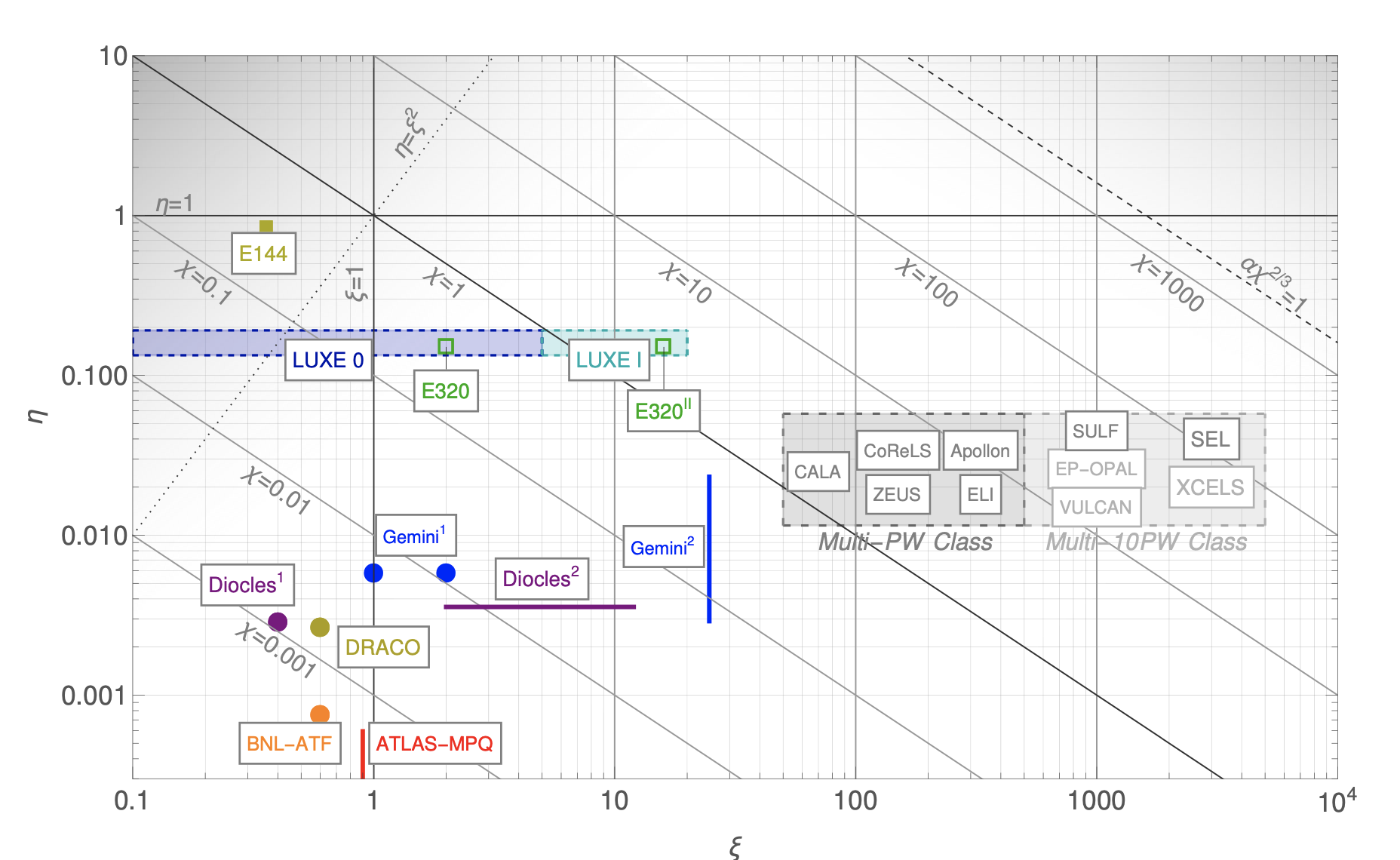}
  \caption{Experimental SFQED laser--particle landscape from~\cite{Fedotov:2022ely}. Dashed Lines and Markers indicate planned experiments while solid lines and markers stand for reported results.}
  \label{fig:SFQED_landscape}
\end{figure*}

In the pioneering E-144 experiment at the Stanford Linear Accelerator Centre (SLAC) in the 1990s, a high-energy electron beam was collided with a neodymium--glass laser pulse, shown in Figure~\ref{fig:E-144}. While achieving field strengths four orders of magnitudes below the Schwinger limit, $\xi\approx0.4$ and $\chi\approx0.27$ were reached. Nonlinear Compton scattering and subsequent nonlinear Breit--Wheeler pair creation were observed, both within the perturbative regime. E-320 at SLAC is the successor experiment with a lower electron beam energy yet more advanced laser technology. In the currently underway experiment E-320 at SLAC's FACET-II, $\xi=7.3$ and $\chi=1.2$ are reachable, enabling the exploration of the nonperturbative regime. At DESY Hamburg, LUXE, a future precision experiment is planned. The landscape of laser--particle experiments is presented in Figure~\ref{fig:SFQED_landscape}. 

\subsection{Crystals, High-Z Fields, Collisions and Astrophysics} 

\textbf{Aligned Crystals:} Electromagnetic fields of nuclear origin can be extended over macroscopic volumes in aligned crystals where atoms are arranged in a regular pattern. This alignment creates axis of symmetry within the crystal, where the associated electric fields of the nuclei add up coherently. Ultra-relativistic positrons or electrons with energy $E_e$ can be \textit{channelled}, meaning they move along the path defined by the aligned nuclei~\cite{DiPiazza:2019vwb}. The coherent summation of the aligned fields experienced by the positron leads to a resulting continuous field strength of order $\mathcal{E} \simeq 10^{13}~\mathrm{V\,m^{-1}}$. The correspondingly reachable quantum parameter is $\chi_{e} \simeq E_e[100~\mathrm{GeV}]$ and typical values for $\xi  \lesssim E_e^{1/2}[100~\mathrm{GeV}]$. At CERN's NA63 experiment, SFQED photon emission and electron--positron pair creation have been observed using aligned crystals and a 200-$\si{\giga\electronvolt}$ lepton beam~\cite{Nielsen:2022bws, Nielsen:2023icv}.

\textbf{High-Z nuclei:} Nuclei with high atomic charge numbers possess a strong Coulomb field near their surface. SFQED effects become significant in heavy elements, like in lead with $Z=82$~\cite{Abramowicz:2021zja} and grow more pronounced for super heavy elements with $Z \geq 92$~\cite{nuclei}. Nonlinear QED processes, such as Delbrück scattering and photon splitting have been observed in nuclei of $Z \gtrsim 80$. Nuclear Schwinger-like vacuum pair creation is estimated to occur when the effective nuclear charge exceeds $Z_{cr}\simeq 173$. Such nuclear fields can be formed transiently in supercritical heavy-ion collisions where the combined charge of two colliding nuclei exceeds this threshold for times on the order of $10^{-23}~\si{\second}$~\cite{Gonoskov:2021hwf}.

Near the surface of heavy nuclei, the electric field strength can approach significant fractions of the Schwinger critical field, making them a natural environment to explore SFQED effects. However, QED effects in such systems are often entangled with nonperturbative nuclear and strong interaction effects. For instance, energy level shifts in high-Z atoms result not only from QED vacuum polarisation and self-energy corrections, but also from finite nuclear size, deformation, and hadronic structure~\cite{Mohr:1998grz}. This overlap complicates the clean identification of QED signatures, making heavy nuclei a challenging environment for isolating SFQED effects~\cite{Abramowicz:2021zja}.

\begin{figure}[tbh]
    \centering
    \includegraphics[width=0.5\linewidth]{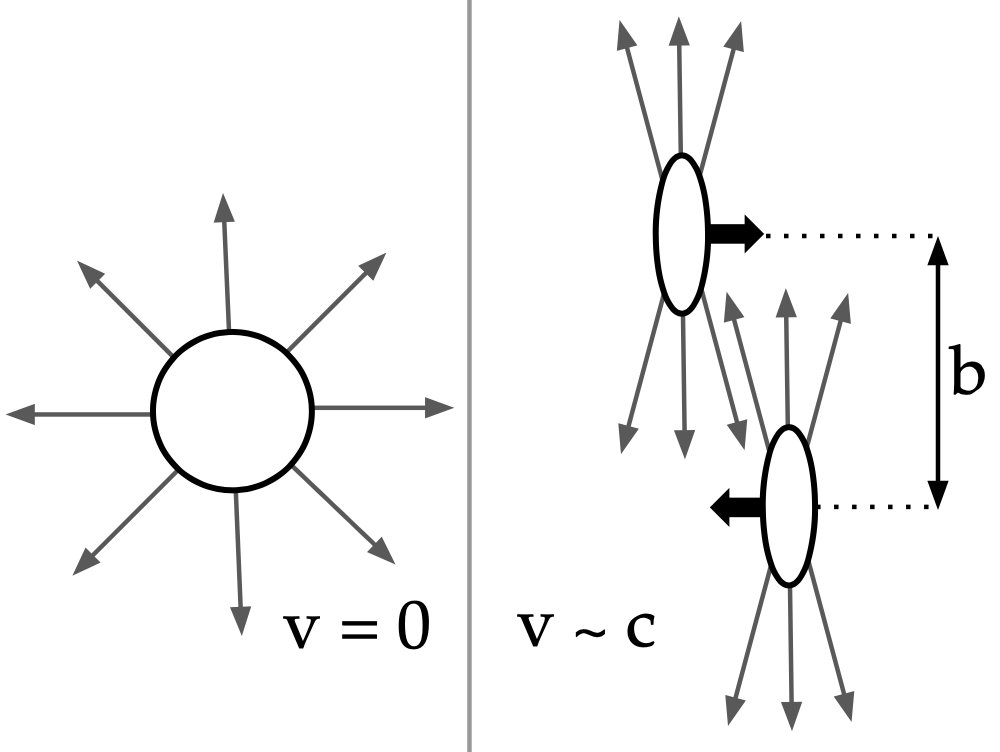}
    \caption{Schematic of Ultra-Peripheral Collisions (UPCs) of heavy ions. }
    \label{fig:ucps}
\end{figure}

\textbf{Ultra-Peripheral Collisions (UPCs):} Ultra-peripheral collisions occur when two heavy ions pass by each other at close distances but without direct nuclear contact, an interaction where field strengths can approach the Schwinger limit in the boosted frame.

This typically happens at impact factor $b$ greater than twice the radius of the ions $b \geq 2r$, and includes that the Coulomb fields are boosted by the relativistic speeds of the particles, depicted in Figure~\ref{fig:ucps}. This interaction is then dominated by QED, because the nuclear contributions are short-range, even for UPCs that involve only protons~\cite{Abramowicz:2021zja}. 

This strategy has been employed by experiments at the RHIC and the LHC using, for example, Pb+Pb scattering.
This has led to the first measurements of both light-by-light scattering $\gamma\gamma \to \gamma\gamma$~\cite{ATLAS:2017fur} and linear Breit--Wheeler pair creation~\cite{CMS:2024tfd, ATLAS:2017fur}. 

While the field strengths can be considered strong for both UPC measurements, leading order perturbation theory applies.
This is because the higher the energy with which the fields of the UPCs collide, the shorter the duration of their overlap. Although the Coulomb fields are strong, the brief duration of their overlap effectively reduces their intensity, and perturbative approaches apply for low-intensity fields~\cite{Fedotov:2022ely}. In other words, even though $\chi_{e,\gamma}$ in these collisions exceeds unity, due to the short duration, $\xi < 1$~\cite{Gonoskov:2021hwf}. Therefore, these are experiments testing QED in strong fields, but not SFQED experiments. 

\textbf{Astrophysics:} Strong magnetic fields surround compact astrophysical objects like neutron stars, pulsars, and black holes. Pulsars are rotating neutron stars with magnetic field strength at the surface of up to $10^9$~T. The rotation of the pulsar magnetic field drives the acceleration of electrons and positrons. The coherent radio emission from pulsars is suspected to be linked to pair cascades resulting from the pulsar's combination of high particle energy and strong magnetic fields~\cite{Gonoskov:2021hwf}.

The Schwinger limit is hypothesised to be surpassed also in magnetars, a type of neutron star believed to possess the strongest magnetic fields observed in the universe with strengths up to $10^{11}$~T~\cite{Gonoskov:2021hwf}. 
Optical polarisation measurements of magnetars have been interpreted as supportive evidence for vacuum birefringence, a SFQED effect occurring in the magnetised vacuum surrounding these objects~\cite{Mignani:2016fwz}.
Additionally, some models consider the possibility that SFQED effects occur near black holes. Powerful gravitational fields near black holes, for example, are hypothesised to convert into strong electromagnetic fields that can surpass the Schwinger limit~\cite{Crumpler:2023toy}.

\section{Summary}

Predictions in QED, like in other quantum field theories, are typically made using perturbation theory. Although the fine-structure constant $\alpha$ runs with energy, its variation is negligible at accessible scales and $\alpha$ remains the small parameter that justifies the perturbative expansion. 
However, in the presence of ultra-strong external fields, where the classical nonlinearity parameter ${\xi \equiv a_0 > 1}$, the interaction with the background field must be treated exactly using the Furry picture. This regime is nonperturbative in the external field but still perturbative in $\alpha$, i.e., nonperturbative at small coupling. At even higher field intensities the perturbative Furry expansion itself may eventually break down when ${\alpha \chi^{2/3} \gtrsim 1}$, as conjectured by Ritus and Narozhny, indicating the onset of fully nonperturbative QED.

Collisions with strong plane-wave fields can be characterised using three key dimensionless parameters: the laser intensity parameter $\xi$, the quantum parameter $\chi$, and the energy parameter $\eta=\chi/\xi$.  
These quantify the classical strength of the background field, the quantum nature of the interaction, and the laser frequency seen in the rest frame of an electron, respectively.
In a plane-wave background, the parameter $\xi$ describes the work done by the field over a Compton wavelength in units of the photon energy and determines how nonlinear the charge--field interaction becomes. The parameters $\eta$  and $\chi$ quantify the importance of linear and nonlinear quantum effects, respectively, such as recoil and pair production.

If $\xi$ and $\chi$ are $\mathcal{O}(1)$, the system enters the nonlinear and nonperturbative regime. Several processes theoretically become accessible that are otherwise suppressed or significantly altered. These include nonlinear (inverse) Compton scattering, where multiple field photons transfer energy to an electron, resulting in the emission of a high-energy photon, and nonlinear Breit--Wheeler pair production, a type of inelastic photon--photon interaction that generates a real electron--positron pair. 
Both processes have been experimentally observed in the nonlinear regime, for instance, in the SLAC E-144 experiment and, more recently, at facilities such as GEMINI~\cite{Poder:2017dpw, Cole:2017zca, Los:2024ysw}, CoReLS~\cite{Mirzaie:2024iey}, or NA63~\cite{Nielsen:2022bws, Nielsen:2023icv}. However, they have not yet been demonstrated in the fully nonperturbative domain, where the Furry expansion is expected to break down.
Advances in laser technology now bring access to higher-intensity regimes where SFQED can be tested with precision at future experiments such as LUXE at DESY Hamburg.

\section*{Acknowledgments}
We thank Tom Blackburn (University of Gothenburg), Anton Ilderton (University of Edinburgh), Ben King (University of Plymouth), and Daniel Seipt (Helmholtz Institut Jena) for leading sessions during the workshop and sharing their expertise. Their openness to experimental perspectives was instrumental in shaping both the workshop and this document. We are grateful for their time, engagement, and support in making strong-field QED more accessible to the experimental community. We also thank the DESY FTX group for supporting the workshop welcome reception. This work was supported by the Swiss National Science Foundation under grants no. 214492 and 230596, and by the PIER workshop funding under ID PWS-2024-17.

\printbibliography

\end{document}